\documentstyle[referee,psfig]{jfm}
\def\be{\begin{equation}}
\def\ee{\end{equation}}
\def\lb{\left[}
\def\rb{\right]}

\begin{document}

\title
[Exact solution of the relativistic Riemann problem]
{The exact solution of the Riemann problem with non-zero
tangential velocities in relativistic hydrodynamics}

\author
[J.A. Pons, J. M$^{\underline{\mbox a}}$ Mart\'{\i} and E. M\"uller]
{Jos\'e A. Pons$^1$,  Jos\'e M$^{\underline{\mbox a}}$ Mart\'{\i}$^1$ 
and Ewald M\"uller$^2$}
              
\affiliation{$^1$Departament d'Astronomia i Astrof\'{\i}sica \\
Universitat de Val\`encia, 46100 Burjassot, Spain \\[\affilskip]
$^2$ Max-Planck-Institut f\"ur Astrophysik, Karl-Schwarzschild-Str. 1,\\
85748 Garching, Germany}

\date{\today}

\maketitle

\begin{abstract}

  We have generalised the {\it exact} solution of the Riemann problem
in special relativistic hydrodynamics (Mart\'{\i} \& M\"uller 1994)
for arbitrary tangential flow velocities. The solution is obtained by
solving the jump conditions across shocks plus an ordinary
differential equation arising from the self-similarity condition along
rarefaction waves, in a similar way as in purely normal flow. The
dependence of the solution on the tangential velocities is analysed,
and the impact of this result on the development of multidimensional
relativistic hydrodynamic codes (of Godunov type) is discussed.

\end{abstract}

\section{Introduction}
\label{s:intro}

  The decay of a discontinuity separating two constant initial states
({\it Riemann problem}) has played a very important role in the
development of numerical hydrodynamic codes in classical (Newtonian)
hydrodynamics after the pioneering work of Godunov (1959). Nowadays,
most modern high-resolution shock-capturing methods (LeVeque 1992) are
based on the exact or approximate solution of Riemann problems between
adjacent numerical cells and the development of efficient Riemann
solvers has become a research field in numerical analysis in its own
(see, e.g., the book of Toro 1997).

  Riemann solvers began to be introduced in numerical relativistic
hydrodynamics at the beginning of the nineties (Mart\'{\i}, Ib\'a\~nez
\& Miralles 1991). Presently, the use of high-resolution
shock-capturing methods based on Riemann solvers is considered as the
best strategy to solve the equations of relativistic hydrodynamics in
nuclear physics (heavy ion collisions) and astrophysics (stellar core
collapse, supernova explosions, extragalactic jets, gamma-ray
bursts). This fact has caused a rapid development of Riemann solvers
for both special and general relativistic hydrodynamics (see, e.g.,
the reviews by Ib\'a\~nez \& Mart\'{\i} 1999, Mart\'{\i} \& M\"uller
1999).

  In a previous paper (Mart\'{\i} \& M\"uller 1994; Paper\,I in what
follows), we derived the analytical solution of the Riemann problem
for an ideal gas in special relativistic hydrodynamics for initial
states where the flow is normal to the initial discontinuity. This
solution has proven to be a useful for (i) the generation of
analytical solutions to test relativistic hydrodynamic codes, and (ii)
the development of numerical hydrodynamic codes based on an exact
Riemann solver (e.g., Mart\'{\i} \& M\"uller 1996; Wen, Panaitescu \&
Laguna 1997).  Because the solution only holds for flows which are
normal to the initial discontinuity, numerical simulations based on
the exact Riemann solver of Mart\'{\i} \& M\"uller (1994) are
restricted to one-dimensional (1D) flows.  Balsara (1994) and Dai \&
Woodward (1997) have circumvented this restriction at the price of
constructing multi-dimensional Godunov schemes using a Riemann solver
based on the two-shock approximation where the rarefaction wave is
treated as a shock wave. An iterative relativistic nonlinear Riemann solver 
that takes into account the effects of nonvanishing tangential velocity 
components has already been implemented by Falle \& Komissarov (1996) for 
Riemann problems involving two strong rarefactions. Finally, we note that the 
effect of the tangential velocity on the properties of simple waves and shocks 
has usually been ignored in relativistic hydrodynamics, because it is always 
possible to choose a reference frame in which the tangential velocity vanishes 
on both sides of the simple wave or shock. In the case of a Riemann problem
this reference frame does not exist. The general solution cannot be
constructed in terms of the solution of the purely normal flow case
and by means of a single Lorentz transformation.

  In the following we derive the exact solution of a Riemann problem
in Minkowski spacetime with arbitrary tangential velocities.  The
solution can be implemented in multidimensional special relativistic
hydrodynamic codes based on directional splitting, because it allows
the computation of the numerical fluxes at every zone interface.
Furthermore, according to recent work by Pons et al. (1998), who solve
the equations of general relativistic hydrodynamics using special
relativistic Riemann solvers, this exact solution can also be
implemented in general relativistic hydrodynamic codes. The exact
solution also allows one to test the accuracy of other approximate
Riemann solvers and codes.

  In this paper we closely follow the structure and notation used in
Paper I, where one can also find the basic references for the theory
of relativistic simple waves and shocks, first discussed by Taub
(1948). Two key references in the theory of relativistic fluids are
the review by Taub (1978) and the book by Anile (1989).  The main idea
behind the solution of a Riemann problem (defined by two constant
initial states, $L$ and $R$, left and right of their common contact
surface) is that the self-similarity of the flow through rarefaction
waves and the Rankine-Hugoniot relations across shocks allow one to
connect the intermediate states $I_*$ ($I = L,R$) with their
corresponding initial states, $I$.  The analytical solution of the
Riemann problem in classical hydrodynamics (see, e.g., Courant and
Friedrichs 1948) rests on the fact that the normal velocity in the
intermediate states, $v^n_{I_*}$, can be written as a function of the
pressure $p_{I_*}$ in that state (and the flow conditions in state
$I$). Thus, once $p_{I_*}$ is known, $v^n_{I_*}$ and all other unknown
state quantities of $I_*$ can be calculated. In order to obtain the
pressure $p_{I_*}$ one uses the jump conditions across the contact
discontinuity, which are given by
\be
p_{L*} = p_{R*} (= p_*)
\ee                                         
\be
v^n_{L*}(p_*) = v^n_{R*}(p_*).
\label{vp0}
\ee
Equation (\ref{vp0}) is an implicit algebraic equation in $p_*$ and
can be solved by means of an iterative method. The function
$v^n_{I*}(p_*)$ is constructed by using the relations across the
corresponding wave connecting the states $I$ and $I*$.

  In the case of relativistic hydrodynamics the same procedure can be
followed, the major difference with classical hydrodynamics stemming
from the role of tangential velocities. While in the classical case
the decay of the initial discontinuity does not depend on the
tangential velocity (which is constant across shock waves and
rarefactions), in relativistic calculations the components of the flow
velocity are coupled through the presence of the Lorentz factor in the
equations. In addition, the specific enthalpy also couples with the
tangential velocities, which becomes important in the
thermodynamically ultrarelativistic regime.

  The structure of the paper is the following. First, in \S2 we
present the equations of relativistic hydrodynamics for a perfect
fluid in 3 spatial dimensions. In \S3 and \S4 we summarise the
properties of the flow across rarefaction waves and shocks,
respectively, and explain how to obtain the pressure and velocities
behind the corresponding waves as a function of the state ahead the
waves. In \S5 we combine the results from the two previous sections to
solve the Riemann problem. Finally, in \S6 we discuss the
implementation of the solution in numerical relativistic
hydrodynamics.  Throughout the paper we will recover the corresponding
purely normal flow expressions whenever it is of interest.

\section{The equations of relativistic hydrodynamics}
\label{s:eqs}

  Let $J^{\mu}$ and $T^{\mu \nu}$ ($\mu, \nu=0,1,2,3$) be the components of the
density current and the energy--momentum tensor of a perfect fluid, 
respectively  
\begin{eqnarray}
J^{\mu} &=& \rho u^{\mu}  \\
T^{\mu \nu} &=& \rho h u^{\mu} u^{\nu} + p \eta^{\mu \nu},
\end{eqnarray}
where $\rho$ denotes the proper rest--mass density, $p$ the pressure, $h = 1 
+ \epsilon + p/\rho$ the specific enthalpy, $\epsilon$ is the specific 
internal energy and $u^{\mu}$ is the four-velocity of the fluid, satisfying 
the normalization condition 
\be 
u^{\mu} u_{\mu} = -1
\label{norm}
\ee
(throughout this paper, we will use the summation convention over repeated 
indices and units in which the speed of light is set to unity). In Cartesian 
coordinates, $x^{\mu}=(t,x,y,z)$, the Minkowski metric tensor $\eta^{\mu \nu}$
is given by 
\be
\eta^{\mu \nu}={\rm diag}(-1,1,1,1).
\ee

  The evolution of a relativistic fluid is determined by the conservation
equation of rest--mass (continuity equation) and energy--momentum
\be
J^{\mu}_{, \mu} = 0 \, ,
\ee
\be
T^{\mu \nu}_{, \mu} = 0 \, 
\ee
where $_{, \mu}$ stands for the partial derivative with respect to coordinate 
$x^{\mu}$. The above system is closed by an equation of state (EOS) which we 
shall assume as given in the form
$p = p(\rho, \epsilon)$.

  The normalization condition of the velocity (\ref{norm}) leads to 
\be
u^{\mu} = W(1,v^x,v^y,v^z)
\ee
where $W$, the Lorentz factor, is
\be
W = (1-v^2)^{-1/2} 
\ee
and 
\be
v^2 = (v^x)^2 + (v^y)^2 + (v^z)^2.
\ee

  The equations of relativistic hydrodynamics admit a conservative formulation
which has been exploited in the last decade to implement high-resolution 
shock-capturing methods. In Minkowski space time the equations in this 
formulation read 
\be 
{\bf U}_{,t} + {\bf F}^{(i)}_{,i} = 0
\label{cl}
\ee
where ${\bf U}$ and ${\bf F}^{(i)}({\bf U})$ ($i=1,2,3$) are, respectively, the
vectors of conserved variables and fluxes 
\be
{\bf U} = (D,S^1,S^2,S^3,\tau)^T
\label{u}
\ee
\be
{\bf F}^{(i)} = (D v^i, S^1 v^i + p \delta^{1i}, S^2 v^i + p \delta^{2i}, 
S^3 v^i + p \delta^{3i},S^i-D v^i)^T.
\label{fu}
\ee
The conserved variables (the rest-mass density, $D$, the momentum density, 
$S^i$, and the energy density $\tau$) are defined in terms of the {\it 
primitive variables},$(\rho, v^i, \epsilon)$, according to
\begin{eqnarray}
D    & = &\rho W             \nonumber \\
S^i  & = &\rho h W^2 v^i               \\
\label{dstau}
\tau & = &\rho h W^2 - p - D. \nonumber
\end{eqnarray}

  System (\ref{cl}) is closed by means of an equation of state that we shall 
assume as given in the form 
\be 
p = p(\rho,\varepsilon).
\ee
The sound speed, $c_s$, is then defined by 
\be
h c_s^2 = \left.\frac{\partial p}{\partial \rho}\right|_s
\ee
where $s$ is the specific entropy.

  In the case of an ideal gas with constant adiabatic exponent, $\gamma$, that 
we have considered in all the tests shown in this paper, the equation of
state is simply
\be
p = (\gamma -1)\rho \varepsilon.
\ee

  The hyperbolic character of the equations of relativistic
hydrodynamics for causal equations of state and their eigenstructure
are well known (e.g., Anile 1989). The complex dependence of the
characteristic fields on the tangential velocity in arbitrary (i.e.,
noncomoving) frames, which is explicitly known (e.g., Donat et
al. 1998), strongly affects the eigensystem and determines the
properties of the waves.

\section{Relation between the normal flow velocity and pressure behind 
         relativistic rarefaction waves}
\label{s:raref}

  Rarefaction waves are simple waves in which the pressure and the
density of a fluid element decreases when crossing them. Choosing the
surface of discontinuity to be normal to the $x$-axis, rarefaction
waves would be self-similar solutions of the flow equations depending
only the combination $\xi=x/t$. Getting rid of all the terms with $y$
and $z$ derivatives in equations (\ref{cl}) and substituting the
derivatives of $x$ and $t$ in terms of the derivatives of $\xi$, the
system reads
\be
(v^x-\xi) \frac{d \rho}{d\xi} +
\{\rho W^2 v^x (v^x - \xi) + \rho \} \frac{dv^x}{d\xi} + 
  \rho W^2 v^y (v^x - \xi)           \frac{dv^y}{d\xi} +
  \rho W^2 v^z (v^x - \xi)           \frac{dv^z}{d\xi} = 0
\label{xi1}
\ee
\be
\rho h W^2 (v^x - \xi) \frac{dv^x}{d\xi} + (1 - v^x \xi) \frac{dp}{d\xi} = 0
\label{xi2}
\ee
\be
\rho h W^2 (v^x - \xi) \frac{dv^y}{d\xi} -  v^y \xi \frac{dp}{d\xi} = 0
\label{xi3}
\ee
\be
\rho h W^2 (v^x - \xi) \frac{dv^z}{d\xi} -  v^z \xi \frac{dp}{d\xi} = 0.
\label{xi4}
\ee
Equation (\ref{xi1}) comes from the equation of continuity. The
remaining equations come from the momentum conservation. Equations
(\ref{xi3})--(\ref{xi4}) reflect the expected result: If the
tangential velocities are zero in the chosen state, no tangential flow
will develop inside the rarefaction. Finally, instead of considering
the conservation of energy, 
we use the conservation of entropy along fluid lines 
(following the same reasoning as in Paper I) which
provides a relation between $dp/d\xi$, $d\rho/d\xi$ and $dh/d\xi$
\be
\frac{dp}{d\xi} = h c_s^2 \frac{d\rho}{d\xi} = \rho \frac{dh}{d\xi}.
\label{xi5}
\ee

  Non-trivial similarity solutions exist only if the determinant of system
(\ref{xi1})-(\ref{xi5}) vanish. This leads to the condition
\begin{eqnarray}
\xi & = & \frac{v^x(1-c_s^2) \pm c_s
\sqrt{(1-v^2)[1 - v^2 c_s^2 - (v^x)^2(1-c_s^2)]}}{1-v^2c_s^2},
\label{xi6} 
\end{eqnarray}
the plus and minus sign corresponding to rarefaction waves propagating
to the left ${\cal R}_{\leftarrow}$ and right ${\cal
R}_{\rightarrow}$, respectively.  It is important to note that the two
solutions for $\xi$ correspond to the maximum and minimum eigenvalues
of the Jacobian matrix associated to the fluxes ${\bf F}^{(x)}({\bf
U})$ (see, e.g., Donat et al. 1998), generalizing the result found for a
vanishing tangential velocity in Paper I (equation (32)).

  After some manipulation, the system (\ref{xi1})-(\ref{xi5}) can be reduced to
just one ordinary differential equation (ODE) and two algebraic conditions
\begin{eqnarray}
\rho h W^2 (v^x-\xi) dv^x + (1-\xi v^x) dp = 0 \label{ode} \label{rar1} \\
h W v^y = \mbox{constant} \label{alg1} \label{rar2} \\
h W v^z = \mbox{constant} \label{alg2} \label{rar3},
\end{eqnarray}
with $\xi$ constrained by (\ref{xi6}). From equations (\ref{rar2}) and
(\ref{rar3}) it follows that $v^y/v^z =$ constant, i.e., the
tangential velocity does not change direction along rarefaction waves
and it is only allowed to change its absolute value. Notice that, in a
kinematical sense, the Newtonian limit ($v^{i} \ll 1$) leads to $W=1$,
but equations (\ref{rar2}) and (\ref{rar3}) do not reduce to the
classical limit $v^{y,z}=$ constant, because the specific enthalpy
still couples the tangential velocities. Thus, even for slow or
moderately relativistic flows ($W \approx 1$),
the Riemann solution presented in this paper must be employed for
thermodynamically relativistic situations ($h > 1$).  The same
result can be deduced from the Rankine-Hugoniot relations for shock
waves (see next section).

  Using (\ref{xi6}), the ODE (\ref{ode}) can be rewritten as
\be
\frac{dv^x}{dp}= \pm \frac{1}{\rho h W^2 c_s}
\frac{1}{\sqrt{1+g(\xi_{\pm},v^x,v^t)}}
\label{ode2}
\ee
where $v^t=\sqrt{(v^y)^2 + (v^z)^2}$ is the absolute value of the tangential
velocity and
\be
g(\xi_{\pm},v^x,v^t)=\frac{(v^t)^2 (\xi_{\pm}^2-1)}{(1-\xi_{\pm} v^x)^2}.
\ee
The sign of $\xi_{\pm}$ corresponds to the sign chosen in
(\ref{xi6}). In the limit of zero tangential velocities, $v^t=0$, the
constants in (\ref{alg1}), (\ref{alg2}) are zero and the function $g$
does not contribute. 
In this limit and in case of an ideal gas EOS one has 
\be
W^2 dv^x = \pm \frac{c_s}{\gamma p} dp = \pm \frac{c_s}{\rho} d\rho,
\ee
recovering expression (30) in Paper I.

  Considering that in a Riemann problem the state ahead of the
rarefaction wave is known, the integration of (\ref{ode2}) allows
one to connect the states ahead ($a$) and behind ($b$) the rarefaction
wave. Moreover, using (\ref{xi6}), the EOS, and the following
relation obtained from the constraint $h W v^t = $ constant,
\be
v^t_b = h_a W_a v_a^t \left\{ \frac{1-(v^x_b)^2}
{h_b^2 + (h_a W_a v_a^t)^2} \right\}^{1/2},
\label{vtraref}
\ee
the ODE can be integrated, the solution being only a function of
$p_b$. This can be stated in compact form as
\be
v^x_b = {\cal R}^a_{\rightleftharpoons}(p_b).
\ee

  Function ${\cal R}^a_{\rightleftharpoons}(p)$ is shown in figure\,1,
for different values of the tangential velocity $v^t$ in state $a$,
the various branches of the curves corresponding to rarefaction waves
propagating towards or away from $a$. Rarefaction waves move towards
(away from) $a$, if the pressure inside the rarefaction is smaller
(larger) than $p_a$. In a Riemann problem the state $a$ is ahead of
the wave and only those branches corresponding to waves propagating
towards $a$ in figure\,1 must be considered. Moreover, one can
discriminate between waves propagating towards the left and right by
taking into account that the initial left (right) state can only be
reached by a wave propagating towards the left (right). The presence
of a tangential velocity in the limiting state restricts the value of
the normal velocity within the rarefaction wave to smaller values as
it is clearly seen from figure\,1. Once the pressure in the post-wave
state has been obtained, the corresponding tangential velocity follows
from (\ref{vtraref}) (see lower panel of figure\,1).

\section{Relation between post-shock flow velocities and pressure
for relativistic shock waves.}

  The Rankine-Hugoniot conditions relate the states on both sides of a shock 
and are based on the continuity of the mass flux and the energy-momentum flux 
across shocks.  Their relativistic version was first obtained by Taub 
(1948) (see also Taub 1978 and K\"onigl 1980).

  If $\Sigma$ is a hyper-surface in Minkowski space time across which $\rho$,
$u^{\mu}$ and $T^{\mu\nu}$ are discontinuous, the relativistic Rankine-Hugoniot
conditions are given by
\be
[\rho u^{\mu}] n_{\mu} = 0 \; ,
\label{rh1}
\ee
\be
[T^{\mu\nu}] n_{\nu} = 0   \; ,
\label{rh2}
\ee
where $n_{\mu}$ is the unit normal to $\Sigma$, and where we have used the
notation
\be
[F] = F_a-F_b,
\ee
$F_a$ and $F_b$ being the boundary values of $F$ on the two sides of
$\Sigma$.

  Considering $\Sigma$ as normal to the $x$-axis, the unitarity of $n_{\nu}$ 
allows one to write it as 
\be
n^{\nu} = W_s (V_s,1,0,0),
\ee 
where $V_s$ is interpreted as the coordinate velocity of the hyper-surface
that defines the position of the shock wave and $W_s$ is the Lorentz
factor of the shock,
\be
W_s = \frac{1}{\sqrt{1-V_s^2}}.
\ee

  Equation (\ref{rh1}) allows one to introduce the invariant mass flux across 
the shock
\be
j \equiv W_s D_a (V_s-v^x_a) = W_s D_b (V_s-v^x_b).
\label{mflux}
\ee
According to our definition, $j$ is positive for shocks propagating to
the right. Note that our convention differs from that of both Landau \& 
Lifshitz (1987) and Courant \& Friedrichs (1948), but is the same as
in Paper I.
                                              
  Next, the Rankine-Hugoniot conditions (\ref{rh1}), (\ref{rh2}) can
be written in terms of the conserved quantities $D$, $S^j$ and $\tau$, 
and of the mass flux as follows
\be
[v^x] = -\frac{j}{W_s} \lb \frac{1}{D} \rb,
\label{vx}  
\ee
\be
[p] = \frac{j}{W_s} \lb \frac{S^x}{D} \rb,  
\label{p}
\ee
\be
\lb \frac{S^y}{D} \rb = 0, 
\label{rhvy} 
\ee
\be
\lb \frac{S^z}{D} \rb = 0, 
\label{rhvz} 
\ee
\be
[v^x p] = \frac{j}{W_s} \lb \frac{\tau}{D} \rb. 
\label{vp}
\ee
We note that in deriving equations (\ref{vx})--(\ref{vp}) we have
made use of the fact that the mass flux is nonzero across a shock.
The conditions across a tangential discontinuity imply continuous
pressure and normal velocity (by setting $j=0$ in equations
(\ref{vx}), (\ref{p}) and (\ref{vp}), and an arbitrary jump in the
tangential velocity.

  Equations~(\ref{rhvy}) and (\ref{rhvz}) imply that the quantity
$hWv^{y,z}$ is constant across a shock wave and, hence, that the
orientation of the tangential velocity does not change. The latter
result also holds for rarefaction waves (see \S3). Equations
(\ref{vx}), (\ref{p}) and (\ref{vp}) are formally identical to the
corresponding equations in Paper\,I (equations (47)-(49)) and can be
manipulated in the same way to obtain $v^x_b$ as a function of $p_b$,
$j$ and $V_s$. Using the relation $S^x=(\tau+p+D)v^x$ and after some
algebra, one finds
\be
v^x_b = \left( h_a W_a v^x_a + \frac{W_s (p_b-p_a)}{j} \right)
\left( h_a W_a + (p_b - p_a) \left(\frac{W_s v^x_a}{j} +
\frac{1}{\rho_a W_a} \right) \right)^{-1}.
\ee
This expression looks like that obtained for vanishing tangential
velocity, but its presence is hidden within the Lorentz factor in
state $a$, $W_a$.  From (\ref{rhvy}) and (\ref{rhvz}) expressions for
$v^y_b$ and $v^z_b$ can be derived
\be
v^{y,z}_b = h_a W_a v^{y,z}_a 
            \left\lbrack 
            \frac{1 - (v^x_b)^2}{h_b^2 + (h_a W_a v^{y,z}_a)^2} 
            \right\rbrack^{1/2} . 
\label{vyzb}
\ee

  The final step is to express $j$ and $V_s$ as a function of the post-shock 
pressure. From the definition of the mass flux we obtain
\begin{equation}
\displaystyle{
V_s^{\pm} = \frac{ \rho_a^2 W_a^2 v^x_a  \pm   
                   |j| \sqrt{j^2 + \rho_a^2 W_a^2 (1 -{v_a^x}^2)}
            }{ \rho_a^2 W_a^2 + j^2 }
}
\label{velshock}
\end{equation}
where $V_s^{+}$ ($V_s^{-}$) corresponds to shocks propagating to the
right (left). 

  The Taub adiabat (Thorne 1973), which relates (only) thermodynamic
quantities on both sides of the shock, and the EOS can be used to
derive the desired expressions. The Taub adiabat, the relativistic
version of the Hugoniot adiabat, is obtained by multiplying
(\ref{rh2}) first by $(h u_{\mu})_a$ and subsequently by $(h
u_{\mu})_b$, and by summing up the resulting expressions. After some
algebra one finds
\be
[ h^2 ] = \left( \frac{h_b}{\rho_b} +
\frac{ h_a}{\rho_a} \right) [p].
\label{taub}
\ee
In the general case, the above non-linear equation must be solved together 
with the EOS to obtain the post-shock enthalpy as a function of 
$p_b$. In the case of ideal gas EOS with constant adiabatic index,
the post-shock density $\rho_b$ can be easily eliminated 
and the Taub adiabat can be rewritten in the form (see Paper I)
\be
h_b^2 \left(1+\frac{(\gamma-1)(p_a-p_b)}{\gamma p_b} \right)
 - \frac{(\gamma -1)(p_a-p_b)}{\gamma p_b} h_b +
\frac{h_a (p_a-p_b)}{\rho_a} - h_a^2 = 0   \; ,
\label{h2}
\ee
which is a quadratic equation for the post-shock enthalpy $h_b$ as a
function of $p_b$. One of the two roots is always negative and must be
discarded as a physical solution.

Next multiplying (\ref{rh2}) by
$n_{\mu}$ and using the definition of the relativistic mass flux
(\ref{mflux}) one obtains
\be
j^2 = \frac{-[p]}{[h/\rho]},
\label{j2}
\ee
which after using the EOS to eliminate $\rho_b$ and inserting the
physical solution of (\ref{taub}) gives the square of the mass
flux $j^2$ as a function of $p_b$.

Using the positive (negative) root
of $j^2$ for shock waves propagating towards the right (left),
equation (\ref{j2}) allows one to obtain the desired relation between
the post-shock normal velocity $v^x_b$ and the post-shock pressure
$p_b$. In a compact way the relation reads
\be
v^x_b = {\cal S}^a_{\rightleftharpoons}(p_b).
\ee

  The function ${\cal S}^a_{\rightleftharpoons}(p)$ is shown in
figure\,2 for several values of the tangential velocity $v^t$ in state
$a$. Its various branches correspond to shock waves propagating
towards or away from $a$. In order to select the relevant branch of
the function ${\cal S}^a_{\rightleftharpoons}(p)$ (figure\,2) the same
argumentation as in the case of rarefaction waves can be used (see
\S3). When $v^x_b$ is known, (\ref{vyzb}) can be used to determine
$v^y$ and $v^z$ in the post-shock state.
 
\section{The solution of the Riemann problem with arbitrary
tangential velocities.}

  The decay of an initial discontinuity gives rise, in general, to
three elementary nonlinear waves (see, e.g., Landau \& Lifshitz
1987). Two of them can be shocks or rarefaction waves, one moving
towards the initial left state and the other towards the initial right
state. Between them, two new states appear, namely $L_*$ and $R_*$,
separated from each other through the third wave, which is a contact
discontinuity moving along with the fluid. Across the contact
discontinuity pressure and normal velocity are constant, while the
density and the tangential velocity exhibits a jump. Accordingly, the
time evolution of a Riemann problem can be represented as:
\be
I\; \rightarrow \;L\;{\cal W}_{\leftarrow}\;L_*\;{\cal C}\;
R_*\;{\cal W}_{\rightarrow}\;R
\ee
where $\cal W$ and $\cal C$ denote a simple wave (shock or rarefaction)
and a contact discontinuity, respectively.  The arrows ($\leftarrow$ /
$\rightarrow$) indicate the direction (left / right) from which fluid
elements enter the corresponding wave.

  As in the Newtonian case, the compressive character of shock waves (density 
and pressure rise across the shock) allows us to discriminate between shocks 
($\cal S$) and rarefaction waves ($\cal R$):
\be
{\cal W}_{\leftarrow\;(\rightarrow)} =
\left\{ \begin{array}{rcl} {\cal R}_{\leftarrow\;(\rightarrow)}
&,& p_b \leq p_a\\
{\cal S}_{\leftarrow\;(\rightarrow)}
&,& p_b > p_a \end{array}
\right.
\label{shandrar}
\ee
where $p$ is the pressure and subscripts $a$ and $b$ denote quantities
ahead and behind the wave. For the Riemann problem $a\equiv L(R)$ and
$b \equiv L_* (R_*)$ for ${\cal W}_{\leftarrow}$ and ${\cal
W}_{\rightarrow}$, respectively. Thus, the possible types of decay of
an initial discontinuity can be reduced to:
\be
(a)\;\; I\; \rightarrow \;L\;{\cal S}_{\leftarrow}\;L_*\;{\cal C}\;
R_*\;{\cal S}_{\rightarrow}\;R \;\;\;\;
p_L < p_{L*} = p_{R*} > p_R
\ee
\be
(b)\;\; I\; \rightarrow \;L\;{\cal S}_{\leftarrow}\;L_*\;{\cal C}\;
R_*\;{\cal R}_{\rightarrow}\;R \;\;\;\;
p_L < p_{L*} = p_{R*} \leq p_R
\ee
\be
(c)\;\; I\; \rightarrow \;L\;{\cal R}_{\leftarrow}\;L_*\;{\cal C}\;
R_*\;{\cal R}_{\rightarrow}\;R \;\;\;\;
p_L \geq p_{L*} = p_{R*} \leq p_R.
\ee
                
  The solution of the Riemann problem consists in finding the
intermediate states, $L_*$ and $R_*$, as well as the positions of the
waves separating the four states (which only depend on $L$, $L_*$,
$R_*$ and $R$). The functions ${\cal W}_{\rightarrow}$ and ${\cal
W}_{\leftarrow}$ allow one to determine the functions $v^x_{R*}(p)$
and $v^x_{L*}(p)$, respectively. The pressure $p_*$ and the flow
velocity $v^x_*$ in the intermediate states are then given by the
condition
\be
v^x_{R*}(p) = v^x_{L*}(p) = v^x_*.
\ee

  When $p_*$ and $v^x_*$ have been obtained the remaining quantities
can be computed. The EOS gives the specific internal energy and the
remaining state variables of the intermediate state $I_*$ can be
calculated using the relations between $I_*$ and the respective
initial state $I$ given through the corresponding wave. In particular,
the tangential velocities can be calculated from (\ref{alg1}) -
(\ref{alg2}) for rarefaction waves and from the Rankine-Hugoniot jump
conditions (\ref{rhvy})-(\ref{rhvz}) for shock waves. Notice that the
solution of the Riemann problem depends on the modulus of $v^t$, but
not on the direction of the tangential velocity.  Figure\,3 shows the
solution of a particular Riemann problem (Sod 1978) for different
values of the tangential velocity $v^y=0, 0.5, 0.9, 0.99$.  The
crossing point of any two lines in the upper panel gives the pressure
and the normal velocity in the intermediate states.  The range of
possible solutions in the ($p, v^x$)--plane is marked by the shaded
region. Whereas the pressure in the intermediate state can take any
value between $p_L$ and $p_R$, the normal flow velocity can be
arbitrarily close to zero in the case of an extremely relativistic
tangential flow. The values of the tangential velocity in the states
$L_*$ and $R_*$ are obtained from the value of the corresponding
functions at $v^x$ in the lower panel of figure\,3.

  To study the influence of tangential velocities on the solution a
Riemann problem, we have calculated the solution of a standard test
involving the propagation of a relativistic blast wave produced by a
large jump in the initial pressure distribution (see Mart\'{\i} \&
M\"uller 1999) for different combinations of $v_L^t$ and $v_R^t$. The
initial data are $p_L=10^3$, $\rho_L=1$, $v^x_L=0$; $p_R=10^-2$,
$\rho_R=1$, $v^x_R=0$ and the 9 possible combinations of $v^t_{L, R} =
0, 0.9, 0.99$. The results are given in figure\,4 and table\,1.

\begin{table}
\begin{center}
\begin{tabular}{|cc|ccccccc|}
\hline\hline
$\quad v^t_L      \quad$ & $\quad v^t_R     \quad$ & 
$\quad \rho_{L*}  \quad$ & $\quad \rho_{R*} \quad$ & 
$\quad p_*        \quad$ & $\quad v^x_*     \quad$ & 
$\quad V_s        \quad$ & $\quad \xi_h     \quad$ & 
$\quad \xi_t      \quad$\\
\hline
0.00 & 0.00 & 9.16e$-2$&1.04e$+1$&1.86e$+1$& 0.960 & 0.987 &$-0.816$&$+0.668$\\
0.00 & 0.90 & 1.51e$-1$&1.46e$+1$&4.28e$+1$& 0.913 & 0.973 &$-0.816$&$+0.379$\\
0.00 & 0.99 & 2.89e$-1$&4.36e$+1$&1.27e$+2$& 0.767 & 0.927 &$-0.816$&$-0.132$\\
\hline
0.90 & 0.00 & 5.83e$-3$&3.44e$+0$&1.89e$-1$& 0.328 & 0.452 &$-0.525$&$+0.308$\\
0.90 & 0.90 & 1.49e$-2$&4.46e$+0$&9.04e$-1$& 0.319 & 0.445 &$-0.525$&$+0.282$\\
0.90 & 0.99 & 5.72e$-2$&7.83e$+0$&8.48e$+0$& 0.292 & 0.484 &$-0.525$&$+0.197$\\
\hline
0.99 & 0.00 & 1.99e$-3$&1.91e$+0$&3.16e$-2$& 0.099 & 0.208 &$-0.196$&$+0.096$\\
0.99 & 0.90 & 3.80e$-3$&2.90e$+0$&9.27e$-2$& 0.098 & 0.153 &$-0.196$&$+0.094$\\
0.99 & 0.99 & 1.29e$-2$&4.29e$+0$&7.06e$-1$& 0.095 & 0.140 &$-0.196$&$+0.085$\\
\hline\hline
\end{tabular}
\end{center}
\caption{Solution of the relativistic Riemann problem at $t=0.4$ with
initial data $p_L=10^3$, $\rho_L=1.0$, $v^x_L=0.0$, $p_R=10^{-2}$,
$\rho_R=1.0$ and $v^x_R=0.0$ for 9 different combinations of
tangential velocities in the left ($v^t_L$) and right ($v^t_R$)
initial state. An ideal EOS with $\gamma=5/3$ was assumed.  The
various quantities in the table are: the density in the intermediate
state left ($\rho_{L*}$) and right ($\rho_{R*}$) of the contact
discontinuity, the pressure in the intermediate state ($p_*$), the
flow speed in the intermediate state ($v^x_*$), the speed of the shock
wave ($V_s$), and the velocities of the head ($\xi_h$) and tail
($\xi_t$) of the rarefaction wave.}
\end{table}

The speeds of the waves propagating to the left and right,
respectively, tend to zero in the limit of $v^t_{L,R} \rightarrow 1$.
This result is easily deduced from (\ref{xi6}) which gives the
velocities of the head ($\xi_h$) and tail ($\xi_t$) of the rarefaction
wave, and from (\ref{velshock}) which gives the shock speed ($V_s$).
These three velocities tend to the normal velocity ahead of the wave
($v^x_a$) when the corresponding tangential velocity ($v^t_a$) tends
to $v^t_{max} = \sqrt{1-(v^x_a)^2}$.

  Although the structure of the solution remains the same (left propagating 
rarefaction wave, right propagating blast wave) the values in the constant 
intermediate states change by a large amount. The results show (see 
figure\,4 and table\,1) that $\xi_h$ remains constant as long as $v^t_L$ is 
constant. The value of $\xi_h$ only depends on the left state, and decreases 
with increasing $v^t_L$. The pressure in the intermediate state, $p_*$, 
increases with increasing $v^t_R$ due to the larger effective inertia of the 
right state (because of the increase of the Lorentz factor of the right state). 
The velocity of the shock is determined by the competing effects of an increased 
$p_*$ and the larger inertia of the right state.  
The shape of the function ${\cal R}^a_{\rightleftharpoons}(p)$ (see
figure\,1) and its dependence on $v^t_L$ shows that both $p_*$ and
$v^x_*$ decrease with increasing $v^t_L$. On the other hand, as a
change in $v^t_L$ does not alter the thermodynamic state, the flow
evolution accross the rarefaction wave is along the same adiabat for
any value of $v^t_L$. This explains why $\rho_L*$ gets smaller with
increasing $v^t_L$ (consistently with the decrease of pressure).
Thus, the net effect of tangential velocities on rarefaction waves is
to evolve further along the adiabat, and to reach smaller intermediate
pressure values.

\section{Conclusions}

  We have obtained the exact solution of the Riemann problem in
special relativistic hydrodynamics with arbitrary tangential
velocities. Unlike in Newtonian hydrodynamics, tangential velocities
are coupled with the rest of variables through the Lorentz factor,
present in all terms in all equations. It strongly affects the
solution, especially for ultra-relativistic tangential flows. 
The specific enthalpy acts as a coupling factor, too. It modifies
the solution in thermodynamically relativistic situations (energy
density and pressure comparable to or larger than the proper rest-mass
density) even in slow flows.

  Our solution has interesting practical applications. First, it can
be used to check the different approximate relativistic Riemann
solvers developed by various authors in the last decade, and to test
multi-dimensional hydrodynamic codes based on directional
splitting. Second, it can be used to construct multi-dimensional
relativistic Godunov type methods. The latter project is currently in
progress and will be reported elsewhere. Finally, using the procedure
described by Pons et al. (1998) the exact solution can be used as a
building block in a general relativistic hydrodynamic code.

  The computational cost of the exact Riemann solver derived above is
comparable to the one presented by Mart\'{\i} \& M\"uller (1994) which
is valid for purely normal flows. Hence, when our exact Riemann solver
is applied to multi-dimensional flow problems, the difference in
efficiency with respect to most linearized Riemann solvers is reduced.

\acknowledgements 
This work has been supported by the Spanish DGICYT grant PB97-1432,
and by the agreement between the Spanish CSIC and the
Max-Planck-Gesellschaft. J.A.P. acknowledges financial support from
the Spanish Ministerio de Educaci\'on y Ciencia (FPI grant).
We acknowledge an unknown referee B for his interesting comments which
helped us to improve the quality of the manuscript.


\newpage

\section*{Figure captions}

{\bf Figure~1}: Loci of states which can be connected with a given state
$a$ by means of relativistic rarefaction waves propagating to the left
(${\cal R}_{\leftarrow}$) and to the right (${\cal R}_{\rightarrow}$)
and moving towards or away from $a$. Upper panel: pressure versus
$v^x$; lower panel: tangential velocity versus $v^x$. Solutions for
different values of the tangential velocity $v^t=0, 0.5, 0.9, 0.953$
correspond to solid, dashed, dashed-dotted and dotted lines,
respectively. The state $a$ is characterized by $p_a=0.6$,
$\rho_a=1.0$, and $v^x_a=-0.3$. An ideal gas EOS with $\gamma=5/3$ was
assumed.
 
{\bf Figure~2}: Loci of states which can be connected with a given state
$a$ by means of relativistic shock waves propagating to the left
(${\cal S}_{\leftarrow}$) and to the right (${\cal S}_{\rightarrow})$
and moving towards or away from $a$. Upper panel: pressure versus
$v^x$; lower panel: tangential velocity versus $v^x$. Solutions for
different values of the tangential velocity $v^t=0, 0.5, 0.8, 0.865$
correspond to solid, dashed, dashed-dotted and dotted lines,
respectively. The state $a$ is characterized by $p_a=0.25$,
$\rho_a=1.0$, and $v^x_a=-0.5$. An ideal gas EOS with $\gamma=5/3$ was
assumed.

{\bf Figure~3}: Graphical solution in the $(p,v^x)$--plane (upper
panel) and in the $(v^t,v^x)$--plane (lower panel) of the relativistic
Riemann problem with initial data $p_L=1.0$, $\rho_L=1.0$,
$v^x_L=0.0$; $p_R=0.1$, $\rho_R=0.125$ and $v^x_R=0.0$ for different
values of the tangential velocity $v^t= 0, 0.5, 0.9, 0.999$,
represented by solid, dashed, dashed-dotted and dotted lines,
respectively. An ideal gas EOS with $\gamma=1.4$ was assumed.  The
crossing point of any two lines in the upper panel gives the pressure
and the normal velocity in the intermediate states. The value of the
tangential velocity in the states $L_*$ and $R_*$ is obtained from the
value of the corresponding functions at $v^x$ in the lower panel, and
$I_0$ gives the solution for vanishing tangential velocity.  The range
of possible solutions is given by the shaded region in the upper
panel.

{\bf Figure~4}: Analytical pressure, density and flow velocity
profiles at $t=0.4$ for the relativistic Riemann problem with initial
data $p_L=10^3$, $\rho_L=1.0$, $v^x_L=0.0$; $p_R=10^{-2}$,
$\rho_R=1.0$ and $v^x_R=0.0$, varying the values of the tangential
velocities. From left to right, $v_{R}^t=0, 0.9, 0.99$ and from top to
bottom $v_{L}^t=0, 0.9, 0.99$. An ideal EOS with $\gamma=5/3$ was
assumed.

\newpage
\begin{figure}
\centerline{\psfig{figure=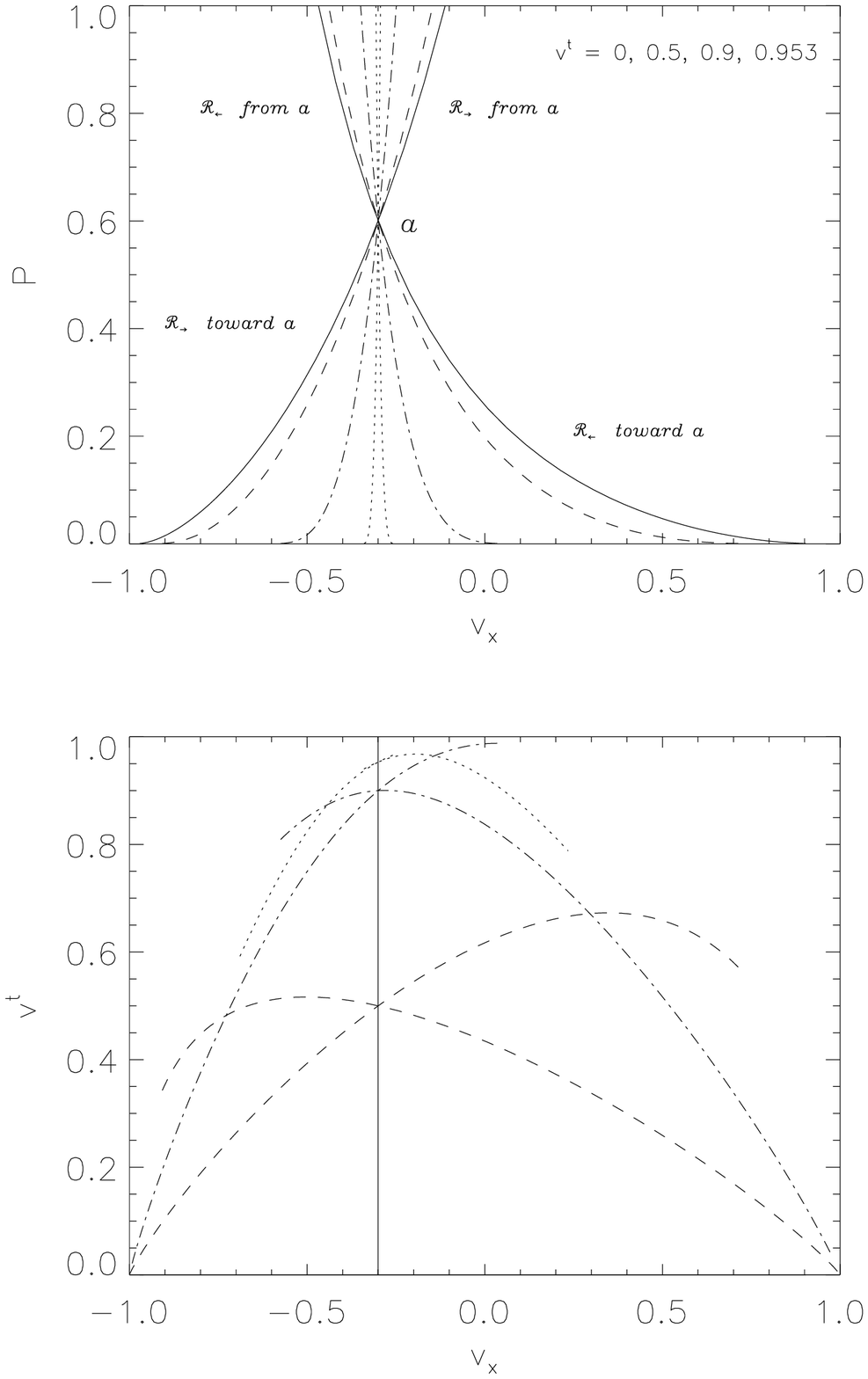,width=14cm}}
\label{fig1}
\end{figure} 

\newpage
\begin{figure}
\centerline{\psfig{figure=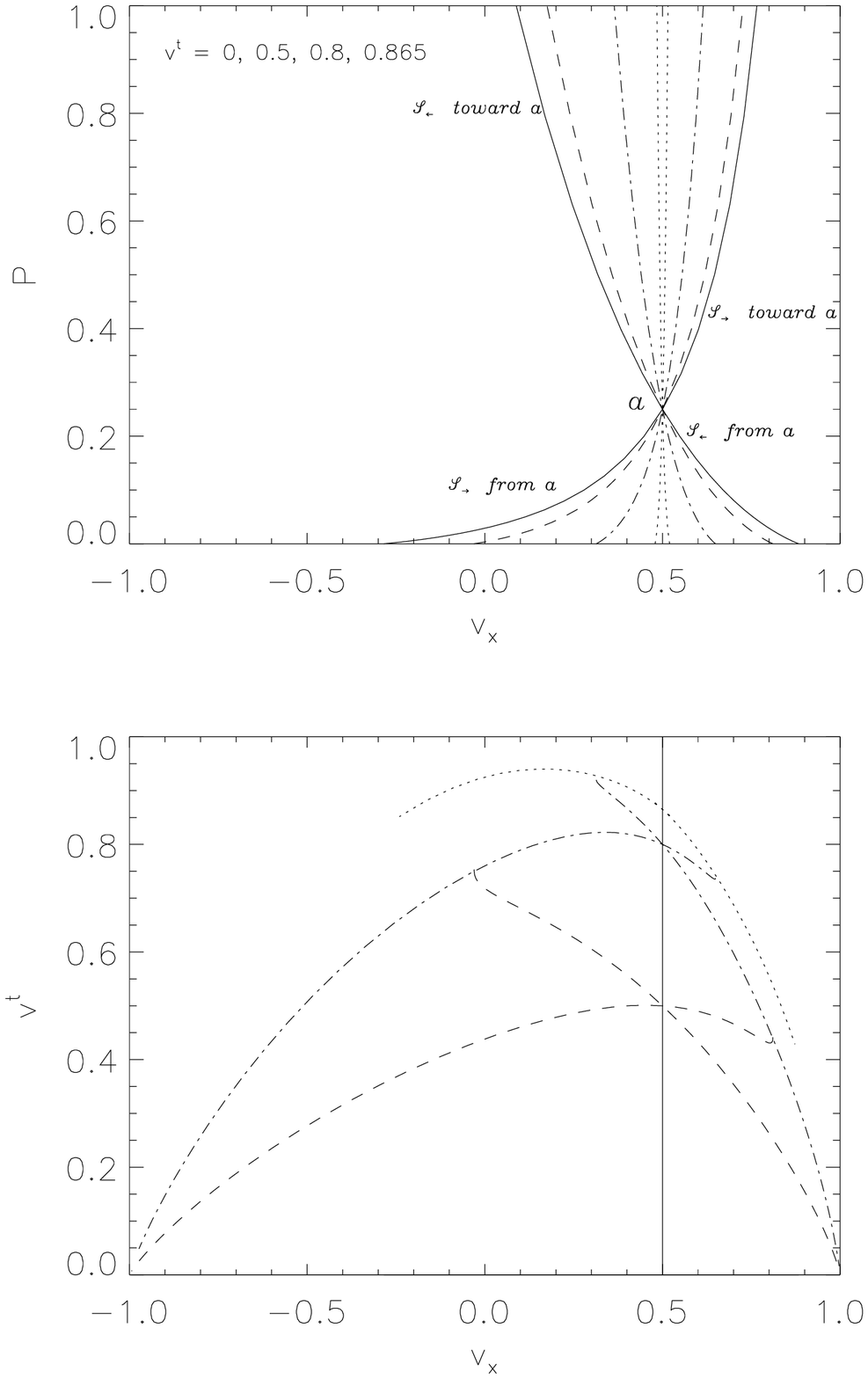,width=14cm}}
\label{fig2}
\end{figure} 

\newpage
\begin{figure}
\centerline{\psfig{figure=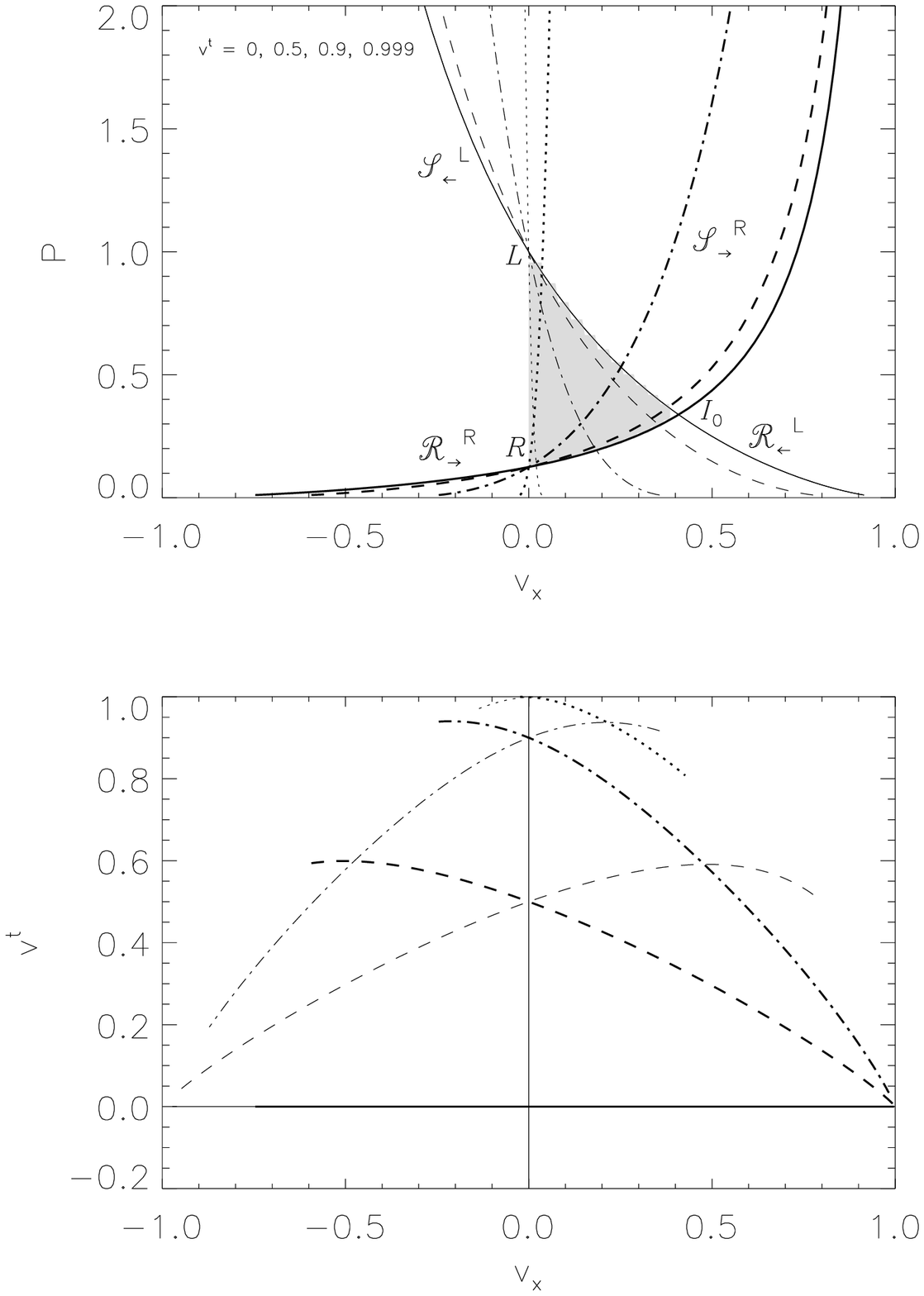,width=14cm}}
\label{fig3}
\end{figure} 

\newpage
\begin{figure}
\centerline{\psfig{figure=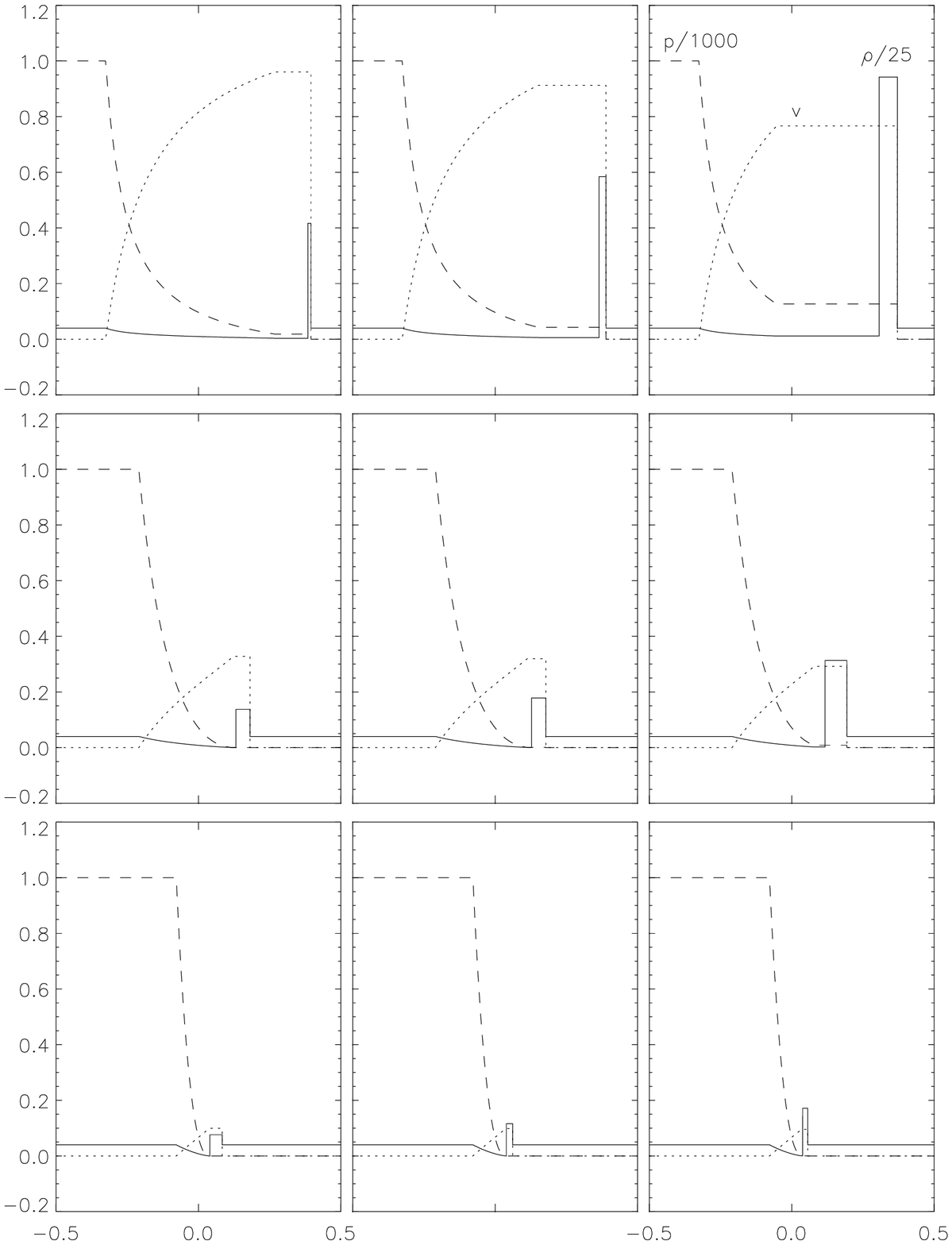,width=14cm}}
\label{fig4}
\end{figure} 

\end{document}